\documentclass[twocolumn,english,apl,aip]{revtex4}
\usepackage[T1]{fontenc}
\usepackage[latin1]{inputenc}
\usepackage{amsmath}
\usepackage{graphicx}
\usepackage{setspace}

\makeatletter

\providecommand{\LyX}{L\kern-.1667em\lower.25em\hbox{Y}\kern-.125emX\@}

\begin{document}

\title{Pseudo Spin Valves Using a (112)-textured D0{$_{22}$} Mn$_{2.3-2.4}$Ga Fixed Layer}

\author{C. L. Zha,$^1$ R. K. Dumas,$^1$  J. Persson,$^1$  S. M. Mohseni,$^1$ J. Nogu\'{e}s,$^{1,3}$ and Johan {\AA}kerman{$^{1,2}$}}

\address{$^1$Material Physics Division, Royal
Institute of Technology (KTH), Electrum 229, 164 40 Stockholm,
Sweden}

\address{$^2$Department of Physics, University of Gothenburg, 412 96
Gothenburg, Sweden}

\address{$^3$Instituci\'{o}Catalana de Recerca i Estudis Avan\c{c}ats  (ICREA) and Centre d'Investigaci\'{o}en
Nanoci\`{e}ncia i Nanotecnologia  (ICN-CSIC), Campus Universitat
Aut\`{o}noma de Barcelona, 08193 Bellaterra, Spain}

\date{\today}

\begin{abstract}

\textbf{Abstract}- We demonstrate pseudo spin valves with a
(112)-textured D0$_{22}$ Mn$_{2.3-2.4}$Ga (MnGa) tilted
magnetization fixed layer and an in-plane CoFe free layer. Single
D0$_{22}$ MnGa films exhibit a small magnetoresistance (MR)
typically observed in metals. In MnGa/Cu/CoFe spin valves a
transition from a negative (-0.08\%) to positive (3.88\%) MR is
realized by introducing a thin spin polarizing CoFe insertion layer
at the MnGa/Cu interface and tailoring the MnGa thickness. Finally,
the exchange coupling between the MnGa and CoFe insertion layer is
studied using a first-order reversal curve (FORC) technique.

\end{abstract}

 \maketitle


\section{INTRODUCTION}
Since the prediction by J. Slonczewski and L. Berger that a spin
polarized current can exert enough torque on a magnetic layer to
significantly affect its magnetization [Slonczewski 1996, Berger
1996], the Spin-Transfer Torque effect (STT) has been intensely
investigated for potential applications in spintronic devices, such
as Spin-Transfer Torque Magnetoresistance Random Access Memory
(STT-MRAM) [Zhu 2008], Spin Torque Oscillators (STO) [Katine 2008],
and domain-wall memory [Parkin 2008]. While such devices are
typically divided into in-plane and perpendicular free or fixed
layer geometries, we recently proposed a so-called Tilted Polarizer
STO where the fixed layer magnetization is tilted out of the film
plane in order to simultaneously achieve zero field operation and
high output power [Zhou 2008, Zhou 2009a]. The tilt angle introduces
an additional degree of freedom, which leads to a surprisingly rich
phase diagram of spin torque switching and precession [Zhou 2009b].

A tilted spin polarizer, with both in-plane and out-of-plane spin
polarization components, can be experimentally achieved using
materials with strong tilted magneto-crystalline anisotropy. We have
previously reported on using (111)-oriented L1$_0$ FePt and FePtCu
with tilted magneto-crystalline anisotropy to fabricate pseudo spin
valves (PSV's) for spin torque devices [Zha 2009a, Zha 2009b, Zha
2009c]. However, L1$_0$ FePt and FePtCu have a number of drawbacks
such as a relatively low spin polarization [Seki 2008], undesirably
high damping factor [Seki 2006], and prohibitive cost due to its
high Pt content. Very recently, D0{$_{22}$} ordered Mn$_{3-x}$Ga
(x=0$\sim$1) was theoretically predicted to be a nearly
half-metallic ferrimagnet with 88\% spin polarization at the Fermi
surface, and was consequently proposed to have great potential for
STT devices [Balke 2007, Winterlik 2008, Wu 2009]. The large
magneto-crystalline anisotropy (K$_{eff}$=$1.2\times10^7$
erg/cm$^3$), the low magnetization ($\leq$M$_s$=250 emu/cm$^3$) and
the expected high degree of spin polarization make Mn$_{3-x}$Ga
ideal as a tilted polarizer [Winterlik 2008, Wu 2009], provided the
appropriate crystalline orientation of the D0{$_{22}$} phase can be
realized.

In this Letter, we report on the successful fabrication of
(112)-textured D0$_{22}$ Mn$_{2.3-2.4}$Ga (MnGa hereafter) thin
films with a tilted magnetization and pseudo spin valves based on
these films. In single D0$_{22}$ MnGa films we observe a small
magnetoresistance (MR) with a parabolic field dependence consistent
with ordinary MR typically observed in all metals.  In MnGa/Cu/CoFe
PSV's a small negative giant magnetoresistance (GMR) is observed
between the MnGa and CoFe layers.  In order to obtain a sizable
positive GMR effect an ultra-thin CoFe layer is inserted at the
MnGa/Cu interface.

\section{EXPERIMENTS}
All film stacks were deposited at room temperature on thermally
oxidized Si substrates using a magnetron sputtering system (AJA ATC
Orion-8) with a base pressure better than $5\times10^{-8}$ Torr.
Deposition of a 6 nm Ta underlayer was followed by MnGa deposition
from a Mn$_{60}$Ga$_{40}$ alloy target. This bilayer was
subsequently annealed $\emph{in-situ}$ at 400$^\circ$C for 35 min to
form the D0{$_{22}$} (112)-textured MnGa phase. For the PSV's a 5 nm
Cu spacer and a 5 nm Co$_{50}$Fe$_{50}$ (CoFe) layer were deposited
after cool-down to room temperature. Finally, a 3 nm Ta capping
layer was deposited on both single MnGa films and PSV's for
oxidation protection. Three different MnGa thicknesses of 15, 25 and
50 nm were employed to fabricate PVS's. The final Mn$_{70}$Ga$_{30}$
film composition was determined using energy dispersive x-ray
spectrometry (EDX). Note that the achieved composition is located in
the 66 to 74 at.\% Mn range where the D0{$_{22}$} phase is expected
to appear [Niida 1996].

Magnetic properties were characterized using a Physical Property
Measurement System (PPMS) equipped with a Vibrating Sample
Magnetometer (VSM) and an Alternating-Gradient Magnetometer (AGM)
with a maximum field of 14 kOe. In addition to standard major
hysteresis loop analysis we employed a first-order reversal curve
(FORC) technique [Davies 2004, Dumas 2007]. First, a family of FORC
curves was measured. Each curve started at a successively more
negative reversal field, H$_R$, and was then measured with an
increasing applied field, H, parallel to the film plane.  Then, a
mixed second-order derivative of the magnetization, M(H, H$_R$), was
used to generate a FORC distribution,
$\rho\equiv$-$\partial^{2}$M(H, H$_R$)/2$\partial$H$\partial$H$_R$,
which was plotted against (H, H$_R$) coordinates on a contour map.
Crystallographic structures were investigated by x-ray diffraction
(XRD) with Cu K$_{\emph{a}}$ radiation in a symmetric scan geometry.
Current-in-plane (CIP) electron transport properties were determined
by a standard 4-point tester with the current orthogonal to the
magnetic field (transverse configuration).
\begin{figure}
\includegraphics[width=0.5\textwidth]{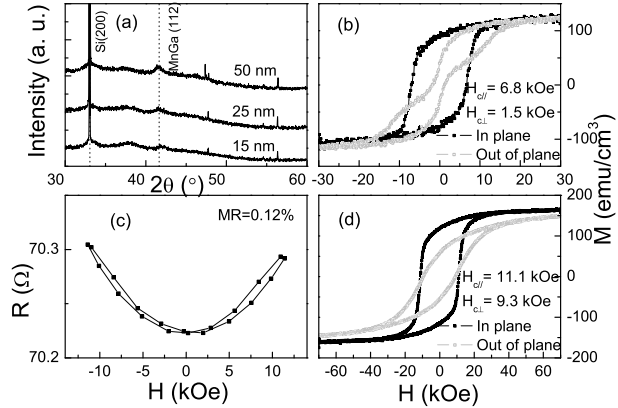}
\caption{\label{fig:FigureXRDVSM} (a) XRD patterns of 15 nm, 25 nm,
50 nm single MnGa films, respectively; (b) in-plane and out-of-plane
VSM hysteresis loops and (c) current-in-plane magnetoresistance
curve of a single 15 nm MnGa film; (d) in-plane and out-of-plane VSM
hysteresis loops of a single 50 nm MnGa film.}
\end{figure}

\section{RESULTS AND DISCUSSION}
Fig. 1 shows structural and magnetic properties of single MnGa
films. Following Ref. [Winterlik 2008], we identify the peaks at
41.36$^\circ$ 
in Fig. 1(a) as the (112) diffraction peak of
the D0$_{22}$ phase of Mn$_{3-x}$Ga. Importantly, with increasing
film thickness we observe enhanced (112) texture as the relative
diffraction intensity becomes stronger. Hysteresis loops of the 15
nm MnGa film in Fig. 1(b) exhibit in-plane coercivity
(H$_{c\parallel}$) of 6.8 kOe and out-of-plane one (H$_{c\bot}$) of
1.5 kOe, as well as a 120 emu/cm$^3$ saturation magnetization
(M$_S$). A 50 nm MnGa (Fig. 1(d)) reveals H$_{c\parallel}$=11.1 kOe,
H$_{c\bot}$= 9.3 kOe, and M$_S$=160 emu/cm$^3$. These values are
consistent with the ferrimagnetic structure of D0$_{22}$ MnGa film
previously reported [Wu 2009]. The squareness ratios of the
out-of-plane and the in-plane loops are less than 1, indicating that
the easy magnetization axis lies in neither the film plane nor along
the normal direction, as expected for highly textured (112)
D0$_{22}$ MnGa films. The improved magnetic properties of the 50 nm
MnGa film are consistent with its enhanced structural properties,
Fig. 1(a). This indicates that the chemical ordering of the
D0$_{22}$ phase of MnGa increases (inducing a higher
magnetocrystalline anisotropy) with thickness under the same
annealing condition. The CIP-MR is found to be 0.12\% at $\pm$14 kOe
for the 15 nm single MnGa film, Fig. 1(c), which we ascribe to
ordinary magnetoresistance of a metal. A relatively large residual
resistance of about 70 Ohm implies a moderate crystallinity,
consistent with the XRD results(Fig. 1(a)).

\begin{figure}
\includegraphics[width=0.5\textwidth]{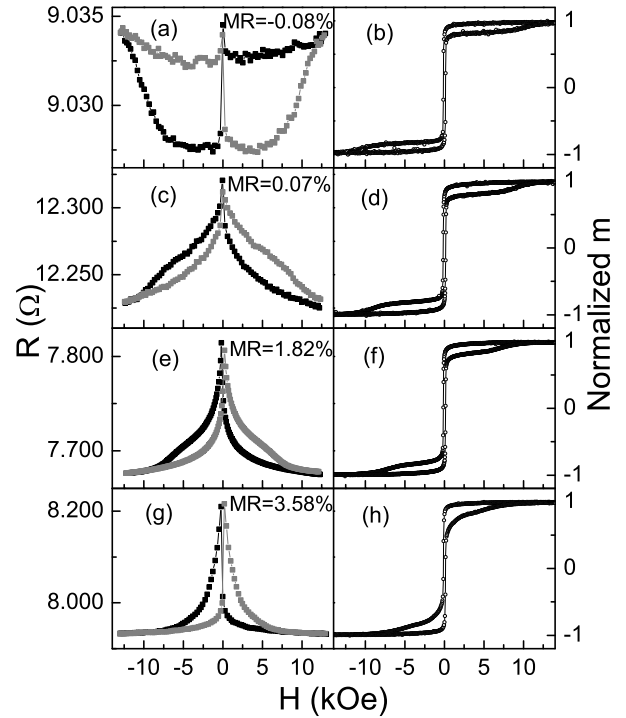}
\caption{\label{fig:FigureXRDVSM}Current-in-plane magnetoresistance
curves (left column) and hysteresis loops (right column) of Ta (6
nm)/MnGa (15 nm)/CoFe (y nm)/Cu (5 nm)/CoFe (5 nm)/Ta (3 nm), (a, b)
y=0; (c, d) y=0.5; (e, f) y=1.0 and (g, h) y=1.5.}
\end{figure}

The magnetotransport and magnetic properties of PSV's of MnGa (15
nm)/CoFe(0, 0.5, 1.0, 1.5 nm)/Cu (5 nm)/CoFe(5 nm) are shown in Fig.
2.  The unique shape of the MR curve in Fig. 2(a), for MnGa/Cu/CoFe,
is due to a combination of effects.  The most prominent contribution
to the MR is a negative GMR between the CoFe and MnGa layers.  Near
positive saturation the two layers are parallel and a relatively
high resistance state is observed.   As the applied field is reduced
(dark filled squares) a sharp drop in resistance near zero field is
observed as the CoFe (5 nm) free layer switches and becomes
anti-parallel to the MnGa layer.  With a further decrease in applied
field the resistance remains nearly constant until roughly -5kOe
where the MnGa layer begins to switch.  Finally, a high resistance
state, corresponding to parallel alignment of the CoFe and MnGa, is
once again achieved at negative saturation.  This situation is
similar to Fe/Cu/GdCo spin valves [Yang 2006] and FeCoGd/AlO/FeCo
tunnel junctions [Bai 2008] where a negative GMR is also observed.
In addition to the dominant negative GMR contribution a small peak
in the resistance near zero field is also observed and is most
likely due to either anisotropic MR in the transverse measurement
geometry or a small positive GMR component.  Finally, a small
parabolic background, most obvious at high fields, is observed due
to the ordinary MR of the MnGa layer, similar to Fig. 1(c).

CoFe usually gives positive bulk and interface spin asymmetry
coefficients with Cu [Li 2002].  Considering, \emph{ab initio}
calculations [Winterlik 2008] of the electronic structure for
D0$_{22}$ Mn$_3$Ga, it is found that the minority (majority) density
of states exhibits a maximum (minimum) at the Fermi energy and the
bulk spin asymmetry coefficient is positive [Tsymbal 1996]. However,
based on the negative GMR found in this PSV, Fig. 2(a), we could not
rule out the possibility of a negative interface spin asymmetry
coefficient at the MnGa/Cu interface.  To obtain a positive
interface spin asymmetry coefficient we employ a thin CoFe insertion
layer at the MnGa/Cu interface. We therefore expect to not only
obtain conventional positive GMR, as anticipated for a PSV, but also
enhancement of the MR [Vouille 1999].

\begin{figure}
\includegraphics[width=0.5\textwidth]{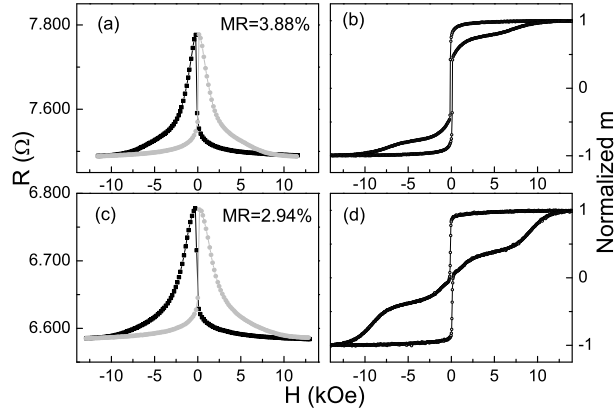}
\caption{\label{fig:AFM}CIP-MR curves (a, c) and in-plane AGM
hysteresis loops (b, d) of Ta (6 nm)/MnGa (25 nm)/CoFe (1.5 nm)/Cu
(5 nm)/CoFe (5 nm)/Ta (3 nm) and Ta (6 nm)/MnGa (50 nm)/CoFe (1.5
nm)/Cu (5 nm)/CoFe (5 nm)/Ta (3 nm), respectively.}
\end{figure}

As shown in Fig. 2(c), when an ultrathin 0.5 nm CoFe layer is
inserted between the spacer and fixed layers, we find a positive MR
of 0.07\%. This suggests that GMR from spin-dependent interface
scattering at the CoFe/Cu/CoFe interfaces is greater than net
negative MR from MnGa alone. The MR is found to increase further as
the thickness of the CoFe insertion layer is increased as shown in
Figs. 2(e) and (g). The shape of the MR loops indicates a two-step
switching for spin valves, which progressively weakens as the CoFe
insertion layer becomes thicker, virtually disappearing for 1.5 nm
layers. On the other hand, the easy magnetization axis is probably
gradually pushed into the film plane by the thicker CoFe.

\begin{figure}
\includegraphics[width=0.4\textwidth]{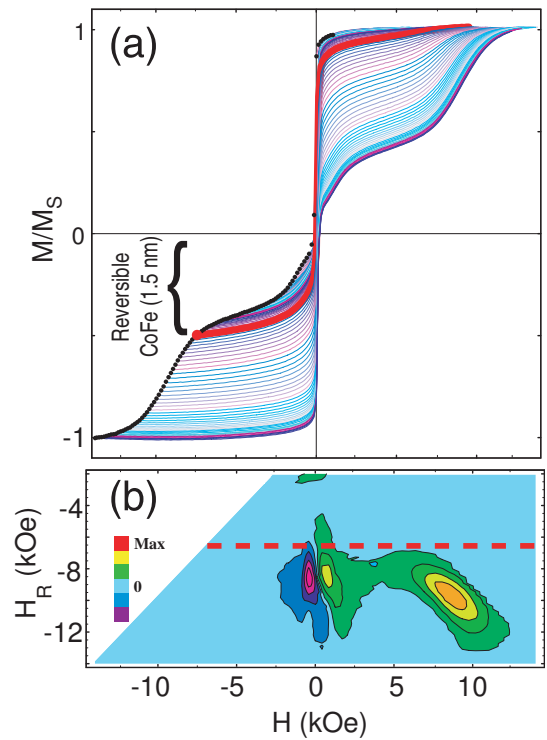}
\caption{\label{fig:AFM} (Color online) (a) A family of FORC curves
and (b) the FORC diagram of MnGa (50 nm)/CoFe (1.5 nm)/Cu (5
nm)/CoFe (5 nm) spin valve. The red horizontal dashed line indicates
the onset of irreversible switching.}
\end{figure}

The hysteresis loops in Figs. 2(b, d, f, h) show a two-stage
switching which leads to the observed GMR effect.  The large
vertical separation between the upper and lower plateaus in the
hysteresis loops is due to the much higher moment (4 times as high)
of the CoFe free layer than the fixed layer. With increasing
thickness of the CoFe insertion layer the switching plateau becomes
gradually smaller and the coercivity of the MnGa fixed layer also
gradually decreases, indicating the switching of the soft CoFe
insertion layer assists the hard MnGa switching due to exchange
interactions.

To further improve current-in-plane magnetoresistance, we explore
the effect of the MnGa thickness on magnetotransport. As shown in
Figs. 3(a) and (c), MR increases to 3.88\% when t$_{MnGa}$=25 nm.
However, when increasing the thickness of MnGa to 50 nm we find the
MR decreases to 2.95\%. This decrease is most likely due to current
shunting through the relatively thick MnGa layer [Zha 2009d]. Figs.
3(b) and (d) exhibit the in-plane magnetic properties for these two
spin valves with t$_{MnGa}$=25 and 50 nm. The separate switching
between the free CoFe layer and the fixed MnGa/CoFe bilayer
corresponds to GMR.

To better understand the interaction between the CoFe insertion
layer and MnGa fixed layer the FORC technique is employed. The
family of measured FORC curves is shown in Fig. 4(a) where black
dots represent the starting point for each FORC. As the reversal
field is decreased a sharp drop in magnetization is found near
H$_R$$\sim$0 kOe which corresponds to the CoFe free layer switching.
This is a highly irreversible process and manifests itself as a very
narrow peak with a large intensity for reversal fields H$_R$$\sim$0
Oe in the FORC distribution.  To highlight the switching of the MnGa
(50 nm)/CoFe (1.5 nm) bilayer alone the FORC distribution is plotted
for H$_R$<-2 kOe in Fig. 4(b). Interestingly, the FORC diagram is
nearly featureless for -6.5 kOe<H$_R$<-2 kOe indicating reversible
switching processes. This region is also highlighted with a bracket
in Fig. 4(a) and is associated with the highly reversible switching
of the CoFe insertion layer. The onset of irreversible switching
occurs for reversal fields H$_R$<-6.5 kOe which is indicated with a
horizontal dashed line in Fig. 4(b) and the red bold FORC in Fig.
4(a). For H$_R$<-6.5 kOe we begin to see peaks in the FORC
distribution that corresponds to irreversible switching of the MnGa
layer as negative saturation is approached. We can interpret the
reversal of the MnGa (50 nm)/CoFe (1.5 nm) bilayer as being that of
a classic bilayer exchange-spring magnet [Davies 2005, Fullerton
1998, Nagahama 1998] where the MnGa and CoFe can be identified as
the hard and soft components, respectively. This exchange spring
interaction explains the lack of a two-step behavior in the MR data
in samples with thick CoFe insertion layers. Essentially, after the
CoFe free layer switches, the CoFe insertion layer begins to
reversibly switch leading to a gradual decrease in the MR as
saturation is approached.

\section{CONCLUSION}

In summary, PSV's using (112)-textured D0$_{22}$ MnGa as a fixed
layer have been demonstrated and a MR up to 3.88\% has been
achieved.  A negative GMR is observed in MnGa/Cu/CoFe spin valves.
However, a negative to positive transition in the MR is realized by
insertion of a thin CoFe layer at the MnGa/Cu interface.  Reversal
of the MnGa/CoFe bilayer has been analyzed and shows exchange-spring
like behavior which explains the lack of a two-step reversal
typically observed in the MR response of spin valves.  These results
are encouraging for future spintronic devices such as STO's where
MnGa will be an advantageous spin polarizer.

\section*{ACKNOWLEDGMENT}

We are grateful to S. Lidin for giving us access to the PPMS. C.Z.
thanks Dr. Ngoc Anh Nguyen Thi for fruitful discussions. Support
from The Swedish Foundation for strategic Research (SSF), The
Swedish Research Council (VR), the G\"oran Gustafsson Foundation,
and the Knut and Alice Wallenberg Foundation is gratefully
acknowledged. J.N. thanks the Wenner Gren Center Foundation, the
Catalan DGR (2009SGR1292) and the Spanish MICINN (MAT2007-66309-C02)
projects for partial financial support. J.{\AA}. is a Royal Swedish
Academy of Sciences Research Fellow supported by a grant from the
Knut and Alice Wallenberg Foundation.

\section*{REFERENCES}

Bai XJ, Du J, Zhang J, You B, Sun L, Zhang W, et al (2008), \emph{J.
Appl. Phys.}, vol. 103, pp. 07F305.

Balke B, Fecher GH, Winterlik J, and Felsera C (2007), \emph{Appl.
Phys. Lett.}, vol. 90, pp. 152504.

Berger L (1996), \emph{Phys. Rev. B}, vol. 54, pp. 9353-9358.

Davies JE, Hellwig O, Fullerton EE, Denbeaux G, Kortright JB, and
Liu K (2004), \emph{Phys. Rev. B}, vol. 70, pp. 224434.

Davies JE, Hellwig O, Fullerton EE, Jiang JS, Bader SD, Zimanyi GT,
and Liu K (2005), \emph{Appl. Phys. Lett.}, vol. 86, pp. 262503

Dumas RK, Li CP, Roshchin IV, Schuller IK, and Liu K (2007),
\emph{Phys. Rev. B}, vol. 75, pp. 134405

Fullerton EE, Jiang JS, Grimsditch M, Sowers CH, and Bader SD
(1998), \emph{Phys. Rev. B}, vol. 58, pp. 12193

Katine JA and Fullerton EE (2008), \emph{J. Magn. Magn. Mater.},
vol. 320, pp. 1217-1226.

Li M, Liao S, and Ju K (2002), US Patent 6683762.

Nagahama T, Mibu K, and Shinjo T (1998), \emph{J. Phys D: Appl.
Phys}, vol. 31, pp. 43

Niida H, Hori T, Onodera H, Yamaguchi Y, and Nakagawa Y (1996),
\emph{J. Appl. Phys.}, vol. 79, pp. 5946-5948.

Parkin SSP, Hayashi M, and Thomas L (2008), \emph{Science}, vol.
320, pp. 190-194.

Seki T, Hasegawa Y, Mitani S, Takahashi S, Imamura H, Maekawa S, et
al (2008), \emph{Nat. Mater.}, vol. 7, pp. 125-129.

Seki T, Mitani S, Yakushiji K, and Takanashi K (2006), \emph{Appl.
Phys. Lett.}, vol. 88, pp. 172504.

Slonczewski JC (1996), \emph{J Magn. Magn. Mater.}, vol. 159, pp.
L1-L7.

Tsymbal EY, and Pettifor DG (1996), \emph{Phys. Rev. B}, vol. 54,
pp. 15314-15329 .

Vouille C, Barthélémy A, Mpondo FE, Fert A, Schroeder PA, Hsu SY, et
al (1999), \emph{Phys. Rev. B}, vol. 60, pp. 6710-6722.

Winterlik J, Balke B, Fecher GH, Alves MCM, Bernardi F, and Morais J
(2008), \emph{Phys. Rev. B}, vol. 77, pp. 054406.

Wu F, Mizukami S, Watanabe D, Naganuma H, Oogane M, Ando Y, et al
(2009), \emph{Appl. Phys. Lett.}, vol. 94, pp. 122503.

Yang DZ, You B, Zhang XX, Gao TR, and Zhou SM, Du J (2006),
\emph{Phys. Rev. B}, vol. 74, pp. 024411.

$^a$Zha CL, Persson J, Bonetti S, Fang YY, and Åkerman J (2009),
\emph{Appl. Phys. Lett.}, vol. 94, pp. 163108.

$^b$Zha CL, Bonetti S, Persson J, Zhou Y, and Åkerman J (2009),
\emph{J. Appl. Phys.}, vol. 105, pp. 07E910.

$^c$Zha CL, Fang YY, Nogués J, and Åkerman J (2009), \emph{J. Appl.
Phys.}, vol. 106, pp. 053909.

$^d$Zha CL, and Åkerman J (2009), \emph{IEEE Trans. Mag.}, vol. 45,
pp. 3491-3494.

Zhou Y, Zha CL, Bonetti S, Persson J, and Åkerman J (2008),
\emph{Appl. Phys. Lett.}, vol. 92, pp. 262508.

$^a$Zhou Y, Zha CL, Bonetti S, Persson J, and Åkerman J (2009),
\emph{J. Appl. Phys.}, vol. 105, pp. 07D116.

$^b$Zhou Y, Bonetti S, Zha CL, and Åkerman J (2009), \emph{New J. of
Phys.}, vol. 11, pp. 103028.

Zhu JG (2008), \emph{Proc. IEEE}, vol. 96, pp. 1786-1798.

\end{document}